\documentclass[useAMS,usenatbib]{mn2e}
\usepackage{epsfig}
\usepackage{graphicx}
\usepackage{aas_macros}
\usepackage{color}
\usepackage{amsmath}
\usepackage{amssymb}
\usepackage{epsfig}
\usepackage{natbib}

\newcommand{\oversim}[2]{\protect{\mbox{\lower0.5ex\vbox{%
   \baselineskip=0pt\lineskip=0.2ex
   \ialign{$\mathsurround=0pt #1\hfil##\hfil$\crcr#2\crcr\sim\crcr}}}}} 
\newcommand{\simgreat}{\mbox{$\,\mathrel{\mathpalette\oversim>}\,$}} 
\newcommand{\simless} {\mbox{$\,\mathrel{\mathpalette\oversim<}\,$}} 

\begin{document}

\title[Was the Progenitor of the Sagittarius Stream a Disc Galaxy?]{
Was the Progenitor of the Sagittarius Stream a Disc Galaxy?}

\author[Pe\~{n}arrubia et al.]{Jorge
  Pe\~{n}arrubia$^1$\thanks{Email: jorpega,vasily,nwe@ast.cam.ac.uk}, Vasily Belokurov$^1$,
  N.W. Evans$^1$, David Mart\'inez-Delgado$^{2,3}$, \newauthor
  Gerard Gilmore$^1$, Mike Irwin$^1$, Martin Niederste-Ostholt$^1$, Daniel B. Zucker$^{4,5}$\\
  $^1$ Institute of Astronomy, University of Cambridge, Madingley Road,
  Cambridge, CB3 0HA, UK \\
  $^2$ Max Planck Institut f\"ur Astronomie, K\"onigstuhl 17, 69117, Heidelberg, Germany\\
  $^3$ Instituto de Astrofisica de Canarias, La Laguna, Spain \\
  $^4$ Department of Physics, Macquarie University, North Ryde, NSW
  2109, Australia\\
  $^5$ Anglo-Australian Observatory, PO Box 296, Epping, NSW 1710, Australia}

 \maketitle  

\begin{abstract} 
We use N-body simulations to explore the possibility that the
  Sagittarius (Sgr) dwarf galaxy was originally a late-type,
  {\em{rotating}} disc galaxy, rather than a non-rotating, pressure-supported dwarf
  spheroidal galaxy, as previously thought. We find that bifurcations
  in the leading tail of the Sgr stream, similar to those detected by
  the SDSS survey, naturally arise in models where the Sgr disc is
  misaligned with respect to the orbital plane.  Moreover, we show
  that the internal rotation of the progenitor may strongly alter the
  location of the leading tail projected on the sky, and thus affect 
  the constraints on the shape of the Milky Way dark matter halo
  that may be derived from modelling the Sgr stream. Our models
  provide a clear, easily-tested prediction: although tidal mass 
  stripping removes a large fraction of the original angular momentum
  in the progenitor dwarf galaxy, the remnant core should still rotate
  with a velocity amplitude $\sim 20$ km\,s$^{-1}$ that could be
  readily detected in future, wide-field kinematic surveys of the Sgr
  dwarf.

\end{abstract} 

%

\section{Introduction}

In spite of extensive theoretical efforts undertaken to reproduce the
characteristics of the Sagittarius (Sgr) stream (e.g. Ibata et
al. 2001; Mart{\'{\i}}nez-Delgado et al. 2004; Helmi 2004; Law et
al. 2005, 2010; Fellhauer et al. 2006), there is presently no
theoretical model that fully explains the wealth of available observational
data. Two aspects of the stream have proved particularly challenging
to model.  First, the leading tail of the Sgr stream is bifurcated,
with both arms exhibiting similar distances, velocities and
metallicity distributions (Yanny et al. 2009; Niederste-Ostholt et al. 2010), which would
appear to refute an earlier model explaining the bifurcation as two
wraps of different ages (Fellhauer et al. 2006), or as independent
streams from different progenitors.  Second, the position on the sky
of the stream suggests that the Milky Way (MW) dark matter halo
interior to the stream has an oblate or perhaps a nearly spherical
shape (Johnston et al. 2005; Fellhauer et al. 2006), while the
heliocentric velocities of stream members support a prolate shape
(Helmi 2004). In a recent work, Law \& Majewski (2010) (hereafter
LM10) have shown that what appears as mutually exclusive results may
in fact signal the possibility of the halo being triaxial in shape, as
Cold Dark Matter (CDM) cosmological models predict (e.g. Kazantzidis et al.2010 and references therein). However, LM10 find that in order to reproduce the
location and velocities of the stream the intermediate axis of the
dark matter halo should be aligned with the spin vector of the MW
disc. This is hard to understand as circular orbits about the
intermediate axis are unstable (e.g. Adams et al. 2007), raising questions as to the formation
of the Galactic disc.  Furthermore, this model does not attempt to
explain the origin of the stream bifurcation.

Interestingly, the chemical composition of Sgr clearly stands out from the
rest of MW dwarf spheroidal galaxies (dSphs). First, strong metallicity
gradients ($-2.3\simless {\rm [Fe/H]} \simless 0.0$), unsual in
other dwarfs, have been reported in the remnant core (Giuffrida et al.2010), which also extends to the tidal
tails (Chou et al. 2007; Monaco et al. 2007; Keller et al. 2010). Also, recent analysis of the chemical
abundances in the core and tails of Sgr suggest that this galaxy
underwent an enrichment history more akin to LMC than to other dSphs (Sbordone et al. 2007; Chou et
al. 2010).  

Motivated by the above results, here we explore the possibility that the Sgr dwarf
was originally a late-type, rotating disc galaxy, rather than a
pressure-supported dSph.
The goal of this paper is to examine how rotation
affects the properties of the associated tidal stream, and to provide model predictions that yield unambiguous tests for this
scenario, rather than reproducing all of the well-documented properties of the Sgr stream in detail.

\begin{figure*}
\includegraphics[width=156mm]{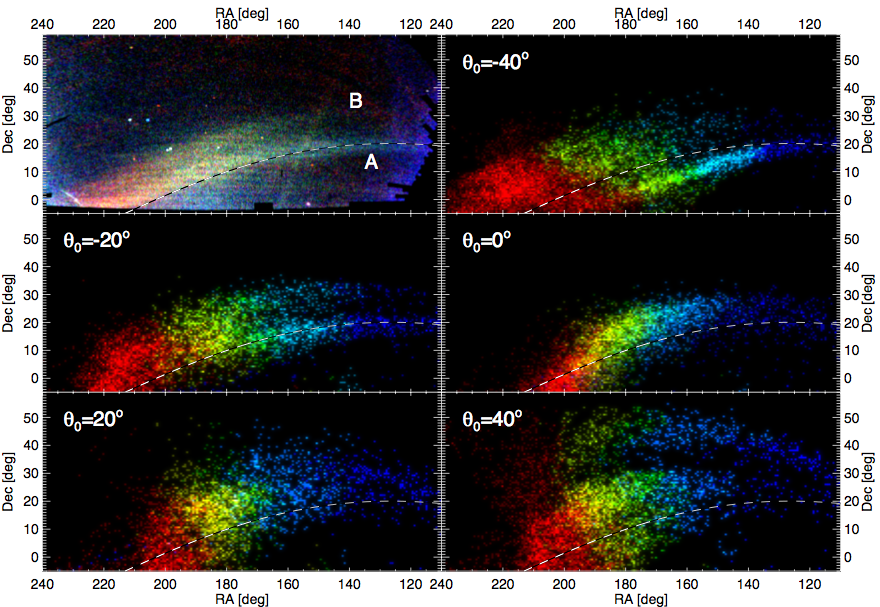}
\caption{ Leading tail particles in the ``Field of Streams'' area
  for different orientations of the Sgr disc with
  respect to its orbital plane (see text). SDSS data are shown in the upper-left panel. Our colour-coding denotes different heliocentric distances (blue, green and red for
  $D<25$, 15--40 and 30--60 kpc, respectively). Dashed lines show a
  projection of the Sgr orbit integrated forward in time. }
\label{fig:fof_6}
\end{figure*}

\section{Numerical Modelling}

\underline{The Galaxy model:} The MW disc is assumed to follow a
Miyamoto-Nagai (1975) model with a mass $M_d=7.5\times 10^{10}
M_\odot$, and radial and vertical scale lengths $a=3.5$ kpc and
$b=0.3$ kpc. The MW bulge follows a Hernquist (1990) profile with a
mass $M_b=1.3\times 10^{10}M_\odot$ and a scale radius $c=1.2$ kpc.
The MW dark matter halo is modelled as a NFW profile with a virial
mass $M_{\rm vir}=10^{12}M_\odot$, virial radius $r_{\rm vir}=258$ kpc
and concentration $c_{\rm vir}=12$ (Klypin et al. 2002). Following
LM10, we assume that the halo is triaxial in shape by introducing
elliptical coordinates with the substitution $r\rightarrow m$, where
$m^2={x^2}/{a^2}+{y^2}/{b^2}+{z^2}/{c^2}$, and $(a,b,c)$ are
dimensionless quantities.  Following the results of LM10, we choose
the density axis-ratios to be $(a,b,c)\simeq (0.61,1.34,1.22)$. The
best-fitting models of LM10 also suggest that the Sun does not sit on
one of the principal axis of the halo, but rather slightly off the x
(i.e. minor)-axis, $(x_\odot,y_\odot)=R_\odot
(-\cos[\lambda],\sin[\lambda])$, where $\lambda\simeq 10^\circ$. We
hold the halo parameters fixed through the evolution of our Sgr models
for simplicity given that tidal streams are barely sensitive to the past evolution of the host potential
(Pe\~narrubia et al. 2006). Our analysis also neglects the effects of
dynamical friction on the orbit of Sgr. However, dynamical friction
is unlikely to introduce a strong orbital decay during the
time-scale of interest (i.e. the last 2--3 Gyr, see Fig. 5 of Pe\~narrubia et al. 2006)

\underline{The orbit of Sgr:} The current position of Sgr and its
line-of-sight velocity in the Galactic standard of rest (GSR) frame
are $(D,l,b)=(25 {\rm kpc}, 5^\circ.6, -14^\circ.2)$ and $v_{\rm
  los}=171$ km s$^{-1}$, respectively (Ibata et al. 1997). To derive
the orbit of Sgr in the above potential, we adopt the proper motion
estimates of Dinescu et al. (2005), who find $(\mu_l\cos
b,\mu_b)=(-2.35\pm 0.20, -2.07 \pm 0.20)$ mas/yr. Using the standard
solar motion about the MW centre 
from Binney \& Merrifield (1998), this translates into a space motion
of $(u,v,w)\simeq (221,-74,203)$ km s$^{-1}$. We use test particles to
integrate its orbit back in time for a few orbital periods in order to
provide initial conditions for our N-body realizations of the Sgr
dwarf. We find that its orbital peri- and apocentres in the above
potential are 15 and 67 kpc, respectively, which implies an orbital
period of 1.04 Gyr. Currently, Sgr has just ($\approx 34$ Myr ago)
gone through its last pericentric passage, and will cross the Milky
Way disc in approximately 28 Myr.

\underline{N-body realizations of Sgr:} We use the algorithm {\sc buildgal} to
construct N-body realizations of late-type spirals composed of a
baryonic disc and a dark matter halo (see
Hernquist 1993 for a detailed description).  The disc profile is
\begin{equation}
\rho_d(R,z)=\frac{m_d}{4\pi R_d^2 z_0}\exp(-R/R_d){\rm sech}^2(z/z_0);
\label{eq:disc}
\end{equation}
where $m_d$ is the disc mass, $z_0$ is the vertical thickness, and $R_d$
the radial scale-length. In our models we
assume that $z_0=0.2 R_d$.  We adopt a non-singular isothermal profile
for the dark matter halo
\begin{equation}
\rho_h(r)=\frac{m_h \alpha}{2\pi^{3/2}r_{\rm cut}}\frac{\exp[-(r/r_{\rm cut})^2]}{r^2 + r_c^2};
\label{eq:halo}
\end{equation}
where $m_h$ is the halo mass, $r_c$ is the core radius and
$\alpha\simeq 1.156$. We find that the properties of the Sgr stream model are not particularly sensitive to the value of the core radius. Here we choose $r_c=0.5R_d$.

We use the results of Nierderste-Ostholt et al. (2010) to crudely
derive a fiducial mass for each of the components of our Sgr N-body
models. These authors estimate that prior to stellar stripping Sgr had
a total luminosity of $L\sim 10^8 L_\odot$, of which only $30$--$50\%$
remains currently bound in the remnant core.  In order to convert
N-body masses into luminosities we adopt a constant stellar
mass-to-light ratio of $\Upsilon_\star =3.5$ (note, however that the
strong metallicity gradients measured throughout the Sgr core may
indicate a large range of $\Upsilon_\star$). Under this choice the
initial Sgr disc mass thus is $m_d\approx 3.5\times
10^8M_\odot$. Adopting a mass-to-light ratio of $\Upsilon=m_h/L\sim
24$, typical for dwarf galaxies with $L\sim 10^8L_\odot$ (Mateo 1998),
we have $m_h=2.4\times 10^9M_\odot$. The total initial mass of our Sgr
model therefore is $M=m_d+m_h=(\Upsilon_\star+\Upsilon)L=2.8\times
10^9M_\odot$.  Also, since we only simulate the most recent history of
the Sgr dwarf, our initial conditions must account for the fact that
the outer halo envelope may have already been lost to tides at the
time when the stellar stream begins to form (see Pe\~narrubia et
al. 2008). To do this we impose a truncation in the dark matter density profile at
$r=r_{\rm cut}=6 R_d$, which roughly corresponds to the tidal radius
of a satellite galaxy with mass $\sim 3\times 10^9 M_\odot$ at a
pericentre $r_{\rm peri}= 15$ kpc.

The only remaining free parameter is the initial scale-length of our
Sgr N-body models, $R_d$. We fix its value by demanding the bound
remnants of the Sgr model to contain $\approx 50$\% of its initial
stellar mass at the final snap-shot of the simulation. Indeed, under
this definition the value of $R_d$ depends on the number of orbital
periods for which we follow the evolution of the Sgr dwarf. Adopting
an integration time of 2.5 orbital periods (see \S\ref{sec:results}
), this
condition yields $R_d\simeq 0.9$ kpc. Under this choice of model
parameters, the peak rotation velocity of Sgr is 43 km s$^{-1}$ at
$R=2.6 R_d=2.34$ kpc, which are indeed typical values for late-type
spirals with luminosities $L\sim 10^8 L_\odot$ (e.g. Swaters et
al. 2009).

\underline{The N-body code:} We follow the evolution of the Sgr N-body
model in the Galaxy potential using {\sc Superbox}, a highly efficient
particle-mesh gravity code (see Fellhauer et al. 2000 for details).
{\sc Superbox} uses three nested grid zones centered on the
highest-density particle cell of the dwarf.  Each grid has $128^3$
cubic cells: (i) the inner grid has a spacing of $dx=3 R_d / 126\simeq
2.4\times 10^{-2} \, R_d$ and is meant to resolve the innermost region
of the satellite galaxy.  (ii) The middle grid extends to cover the
whole dwarf, with spacing $10 R_d/ 126$. (iii) The outermost grid
extends out to $50\times r_{\rm vir}$ and is meant to follow particles
that are stripped from the dwarf. We choose a constant time-step
$\Delta t=0.9$ Myr, which leads to a total energy conservation better
than 1\% after the dwarf models are evolved for a Hubble time in
isolation. This shows that the evolution of the N-body model is free
from artefacts induced by finite spatial and time resolutions.

\begin{figure*}
\includegraphics[width=140mm]{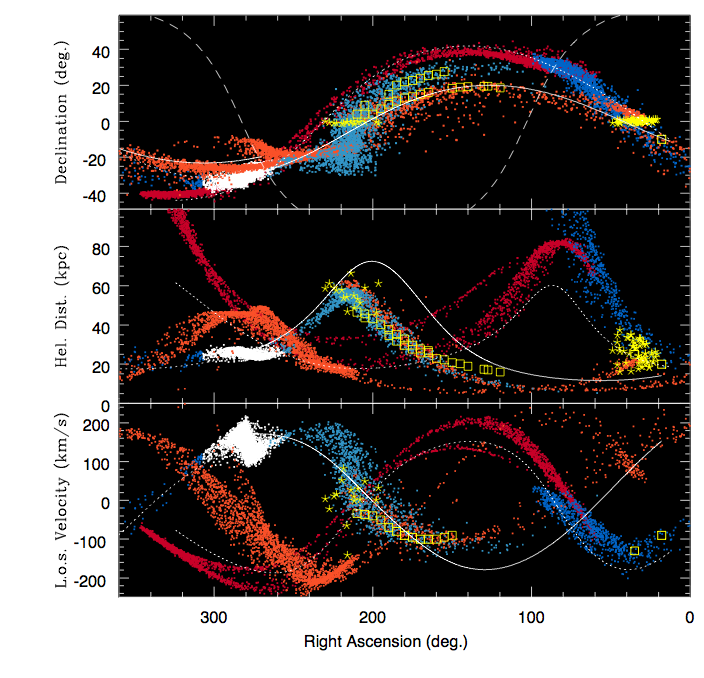}
\caption{ Projected location ({\it upper panel}), heliocentric
  distance ({\it middle panel}) and line-of-sight velocity ({\it lower
    panel}) in the GSR frame of the leading/trailing (light/dark
  colours) tail particles of a Sgr stream model with initial
  orientation $\theta_0=-20^\circ$. Yellow stars and open squares
  denote RR Lyr (Vivas et al. 2005, Watkins et al. 2009) and SDSS
  detections of Sgr (Belokurov et al. 2006). White particles 
  show the remnant core of the Sgr dwarf (see also
  Fig~\ref{fig:core}). The dashed line marks the position of the Galactic plane. For ease of reference we also show its orbit integrated forward (solid lines) and backward (dotted
  lines) in time.  Note that the bifurcation of the leading arm arises
  from material stripped at different pericentric passages: blue at
  the penultimate pericentre passage ($\simeq 1$ Gyr ago) and red at
  the ante-penultimate one ($\simeq 2$ Gyr ago).}
\label{fig:model}
\end{figure*}

\section{Results}\label{sec:results}

The angle subtended by the internal angular momentum vector of Sgr and
that of its orbit about the MW, i.e. $\theta_0\equiv {\rm
  acos}[\hat{{\bf J}}_{\rm int}\cdot \hat{{\bf J}}_{\rm orb}]$ is a
free parameter in our study. 
The orientation of the Sgr disc is set at the start of the simulation by rotating ${{\bf J}}_{\rm int}$ an angle $\theta_0$ about the instantaneous line of the orbit.
Under this definition, $\theta_0=0$
denotes a model where the spin vector of the Sgr disc and the normal
vector of the orbital plane are perfectly aligned, whilst $\theta_0>0$
and $\theta_0<0$ respectively indicate models where Sgr rotates in a
prograde and retrograde motion with respect to its Galactic orbit. 

Fig.~\ref{fig:fof_6} shows the projection on the ``Field of streams''
area of the sky (Belokurov et al. 2006) of Sgr stellar debris for
different disc orientations after integrating our N-body models for
2.5 orbital periods. For simplicity, we show only particles that
belong to the leading tail of the Sgr stream, given that the trailing
tail has not been detected yet in the Northern Galactic Hemisphere.
This Figure illustrates a few interesting points. The first is during
the stripping process a fraction of the internal angular momentum of
Sgr transfers to the stream. As a result, the stream tail(s) do not
trace the progenitor's orbit (dashed lines in
Fig.~\ref{fig:fof_6}). In practical terms, our results suggest that
internal rotation in the progenitor of the Sgr stream may have to be
taken into account when inferring the shape of the MW halo from
fitting the position and velocity of stream pieces. A second
interesting point is that bifurcations in the leading tail of the Sgr
stream naturally appear in this area of the sky if the spin vector and
the normal vector of the orbital plane are misaligned. The separation
between the bifurcated arms clearly becomes more prominent as the
value of $|\theta_0|$ increases. 
Interestingly, by colour-coding the
stellar particles according to their heliocentric distances, we can
appreciate that in all models both arms show a very similar gradient
throughout the sky, in concordance with observational data (Belokurov
et al. 2006; Niederste-Ostholt et al. 2010). Hence, internal rotation in the Sgr dwarf mainly affects the apparent precesion of the stream plane on the sky.

The model with
$\theta_0=-20^\circ$ has a striking resemblance with the Field of Streams. It also provides a reasonable match to most of the existing observational constraints. 
Fig.~\ref{fig:model} shows the projected location ({\it upper panel}),
heliocentric distance ({\it middle panel}) and line-of-sight velocity
({\it lower panel}) of the leading (open symbols) and trailing (closed
symbols) tail particles of a Sgr stream model with initial orientation
$\theta_0=-20^\circ$.  By colour-coding the particles according to the
time at which they are stripped, we can appreciate that the
bifurcation arises from material lost at consecutive pericentric
passages. In particular, the southern, more prominent tail of the
leading arm (stream A) corresponds to stars that were lost at the
third most recent pericentric interaction, i.e. $\simeq 2$ Gyr ago
(coloured in red), whilst the fainter northern tail (stream B) is more
recent and is composed of stars that became unbound at the penultimate
pericentre, i.e. $\simeq 1$ Gyr ago (in blue).  This result clearly
implies that the minimum age of the Sgr stream is two orbital periods,
which justifies the choice of integration time for our N-body models.

This model correctly reproduces the distances and velocities measured
along streams A and B, as well as the recent detections of the
trailing tail in the southern hemisphere. Puzzlingly, although also
predicted by our model, the presence of the trailing tail in the Field
of Streams has thus far eluded detection. This may be explained by the
drop of stars expected beyond the apocentre of the trailing tail,
i.e. R.A.$\simgreat 90^\circ$, as well as the sharp increase in distance 
at R.A.$\simless 120^\circ$, which may both conspire to put the surface brightness and 
the turn-off apparent magnitude of the stream beyond the detection threshold of current photometric
surveys.

\begin{figure}
  \includegraphics[width=84mm]{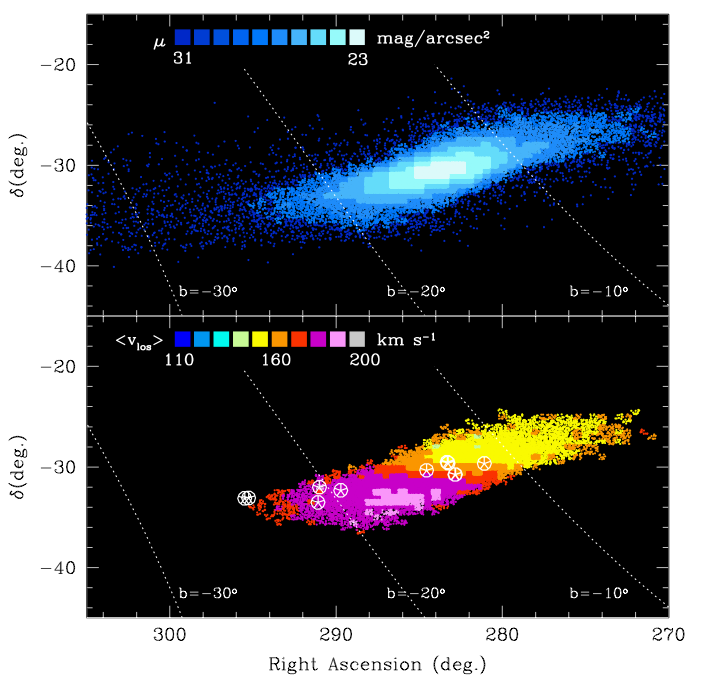}
  \caption{The remnant core of the Sgr dwarf model shown in
    Fig.~\ref{fig:model}. Particles are colour-coded according to the
    dwarf surface brightness at their location ({\it upper panel}) and
    mean line-of-sight velocity ({\it lower panel}). Symbols denote
    the location of the fields spectroscopically surveyed by Ibata et
    al. (1997). Note that our models predict that the remnant core still
    rotates with a velocity amplitude $\sim 20$ km s$^{-1}$.}
\label{fig:core}
\end{figure}

Fig.~\ref{fig:core} suggests that further clues on the nature of the
Sgr dwarf may be gained by studying its remnant core.  In the upper
panel, we show the projection onto the sky of the Sgr dwarf model
shown in Fig.~\ref{fig:model}. Particles are colour-coded according to
the dwarf surface brightness at their location. The core morphology
agrees well with that derived from the 2MASS survey (see e.g. Fig. 4
of Majewski et al. 2003), although its central surface brightness is
somewhat brighter than estimated by these authors ($\mu_0\simeq
24.8$mag/''$^2$). Mismatches in surface brightness, however, may be a
consequence of adopting a constant stellar mass-to-light ratio for our
N-body particles.  The lower panel of Fig.~\ref{fig:core} shows that,
although tidal mass stripping removes a large fraction of the original
angular momentum in the progenitor Sgr disc, the remnant core is
predicted to rotate with a velocity amplitude $\sim 20$ km s$^{-1}$,
which translates into a net velocity difference of $\sim 40$ km
s$^{-1}$ between stars at declinations above and below $\delta \sim
-30^\circ$. Although not shown here, the central line of sight
velocity dispersion is $\simeq 12$ km s$^{-1}$ , in reasonable
agreement with the estimates of Bellazzini et al. (2008), who find a
uniformly flat profile at $\sigma\simeq 10$ km s$^{-1}$ over the
central $0'\le r\le 9'$ range.

The only available kinematic survey throughout the Sgr dwarf dates
back to Ibata et al. (1997). These authors measured a line-of-sight
velocity between 160--170 km s$^{-1}$ in most of their pencil-beam
fields, hence with little or no evidence of net rotation. Although in
these locations our model accurately reproduces their velocity
measurements, Fig.~\ref{fig:core} suggests that Ibata et al. fields
were too sparsely sampled to pick up the rotation signal predicted by
our models.

\section{Conclusions}

Here, we explored the possibility that the Sgr was originally a
late-type, rotating disc galaxy. We show that bifurcations in the
leading tail of the Sgr stream, similar to those detected by the SDSS
survey, naturally arise in models where the disc is misaligned with
respect to the orbital plane.  This occurs because material is
primarily stripped at pericentric passages, and successive passages
occur at different orientations of the Sgr disc.  Together with
observed metallicity patterns that are atypical for dwarf spheroidal
galaxies, this suggests that the Sgr dwarf may have originally been a
galaxy akin to late-type spirals with a peak rotation of $\simgreat 45$
km\,s$^{-1}$.  We also find that internal rotation alters the position
of the stream with respect the Sgr orbit, which might indirectly
affect any constraint on the shape of the MW halo derived from
pressure-supported dwarf models.

Fortunately, before we embark upon more complex modelling of the Sgr
stream, there is a clear-cut way to test whether the Sgr dwarf was
indeed a rotating galaxy. Although tidal stripping efficiently removes angular momentum from the progenitor dwarf, we find that the remnant core should still
rotate with a velocity amplitude close to $\sim 20$ km\,s$^{-1}$ given the current estimates of the fraction of light residing in the tidal stream.   
Validating
this prediction is feasible with existing instruments and will shed
light on the true origin of the Sgr dwarf, as well as on the shape of
the MW dark matter halo.

\vskip0.5cm
JP and MNO acknowledge financial support from the Science and
Technology Facilities Council of the United Kingdom, whilst VB thanks
for the Royal Society for financial support. This work has benefited greatly from discussions with D. Lynden-Bell. We thank the referee, D. Law, for his thorough read of this manuscript and his useful comments.

{} 
\end{document}